# Giant nonlinear conduction and thyristor-like negative derivative resistance in BaIrO$_3$ single crystals

Tomohito NAKANO* and Ichiro TERASAKI**

*Department of Applied Physics, Waseda University, 3-4-1 Okubo Shinjuku-ku Tokyo 169-8555, Japan*

*CREST, Japan Science and Technology Agency, Kawaguchi, Saitama 332-0012, Japan*

Abstract

We synthesized single-crystalline samples of monoclinic BaIrO$_3$ using a molten flux method, and measured their magnetization, resistivity, Seebeck coefficient and nonlinear voltage-current characteristics. The magnetization rapidly increases below a ferromagnetic transition temperature $T_C$ of 180 K, where the resistivity concomitantly shows a hump-type anomaly, followed by a sharp increase below 30 K. The Seebeck coefficient suddenly increases below $T_C$, and shows linear temperature dependence below 50 K. A most striking feature of this compound is that the anomalously giant nonlinear conduction is observed below 30 K, where a small current density of 20 A/cm$^2$ dramatically suppresses the sharp increase in resistivity to induce a metallic conduction down to 4 K.

*Present address: Department of Applied Physics, Waseda University, 3-4-1 Okubo Shinjuku-ku Tokyo, JAPAN 169-8555. :Tel./Fax: +81-3-5286-3854

:E-mail address: nakano@htsc.sci.waseda.ac.jp

**Present address: Department of applied physics, Waseda University, 3-4-1 Okubo Shinjuku-ku Tokyo, JAPAN 169-8555 Tel./Fax: +81-3-5286-3854

:E-mail address: terra@waseda.jp

An important purpose of condensed matter physics is to understand emergence arising from numerous degrees of freedom. Prime examples are phase transitions and collective excitations: A ferromagnetic transition induces a collective mode of spins known as a magnon, whereas superconductivity lets an electric current flow persistently by spatially changing the phase of the superconducting order parameter. Recent studies of emergence have revealed the importance of competing and/or cooperating orders, in which novel phenomena such as intrinsic inhomogeneity [1-3], quantum criticality [2-6] and multiferroics [7-9] are proposed theoretically, and are discovered in real materials.

In the context of competing/cooperating orders, the barium iridate $BaIrO_3$ is of extreme interest, which was synthesized first in 1960s by Donohue *et al*. [10]. It exhibits a charge density wave (CDW) transition accompanying a ferromagnetic transition at the same temperature of $T_C$=180 K [11-13]. The CDW transition was characterized by a hump-type resistivity anomaly and a gap-like optical spectrum [11]. The Seebeck coefficient [14], photoemission [15], band calculation [16] and magnetoresistance [17] suggest that the gap is opened in the density of states below $T_C$. However, a satellite scattering in the X-ray diffraction pattern is not yet reported at present, and thus it needs more discussion whether the gap below $T_C$ could be associated with a conventional CDW or not.

The crystal structure of $BaIrO_3$ is monoclinic (space group C2/m) [18, 19], and contains zigzag chains of the $Ir_3O_{12}$ trimer with the face shared $IrO_6$ octahedra along the c-axis as shown in Fig 1. Each trimer is linked to form a corner-shared network along the c axis, and a two-dimensional network perpendicular to the c axis as well. The band calculation revealed a fairly dispersive band along the ab plane as well as along the c axis [16].

We have studied competing/coexisting orders in strongly correlated systems. We found a novel transition from localized to itinerant antiferromagnetism in the heavy-fermion compound

Ce(Ru$_{0.85}$Rh$_{0.15}$)$_2$Si$_2$ [20-22], and discovered a giant nonlinear conduction in the organic conductor θ-(BEDT-TTF)$_2$CsM(SCN)$_4$ (M=Co, Zn) [23,24] where two different charge-ordered domains coexist [25]. In this paper, we report that BaIrO$_3$ single crystals exhibit giant nonlinear conduction as a function of pulsed external current, and shows a negative derivative resistance from 2 to 6 A/cm$^2$ at 4.2 K. This would belong to a new class of collective excitation in solids to be associated with the giant nonlinear conduction in the organic conductor.

Single-crystalline samples of BaIrO$_3$ with a typical dimension of 1 × 0.5 × 0.2 mm$^3$ were synthesized by a molten flux method using BaCl$_2$ as solvent. Figure 2(a) shows the X-ray diffraction (XRD) on a powdered sample prepared by grinding an as-grown BaIrO$_3$ single crystal, indicating that the sample was fully indexed with the monoclinic structure. The orientation of the single crystals was determined by the Weisenberg method using an R-AXIS RAPID (RIGAKU) with Mo $k\alpha$ radiation. A typical x-ray diffraction image of a single-crystalline sample is shown in Fig 2 (b), where clear spots warrant the crystal quality.

The magnetization was measured in a magnetic field of 1 kOe with a commercial SQUID magnetometer (Quantum Design). The resistivity was measured by a standard four-probe method in a liquid He cryostat. The voltage-current characteristics were measured on the same sample with a pulsed current of 10 ms using a pulse current source (Keithley 6221) from 4.2 to 120 K. The Seebeck coefficient was measured using a steady-state technique from 4.2 to 300 K in a liquid He cryostat.

The magnetization of a BaIrO$_3$ single crystal in a magnetic field of 1 kOe parallel and perpendicular to the c-axis are shown in Fig. 3(a), exhibiting a rapid increase below $T_C$ = 180 K (marked by a vertical arrow in this figure). This is direct evidence for the ferromagnetic transition in this compound. Below about 100 K, the magnetization perpendicular to the c-axis is slightly smaller than that parallel to the c-axis, which is consistent with the previous results [11-13]. We should note that there is no trance of phase transitions other than at 180 K, which is consistent with the specific heat data [14].

Figure 3 (b) shows the resistivity $\rho_{//}$ and $\rho_{\perp}$ for an external current density $J$ (= 0.01 A/cm$^2$) parallel and perpendicular to the c-axis, respectively. $\rho_{//}$ slightly increases with decreasing temperature from room temperature down to $T_C$ to show nonmetallic conduction. Then it shows a hump-type anomaly just below $T_C$, and saturates at around 120 K. The hump implies that a part of the Fermi surface is gapped below $T_C$. Similar anomalies have been observed in NbSe$_3$ [26], Cr [27] and Ce(Ru$_{0.85}$Rh$_{0.15}$)$_2$Si$_2$ [28]. Below 120 K, $\rho_{//}$ decreases with decreasing temperature like a metal. We think that the electric conduction is essentially metallic below $T_C$, although we cannot see a clear metallic conduction just below $T_C$ where the gap opening dominates $\rho_{//}$. We may thus say that a "nonmetal-to-metal" transition occurs at the ferromagnetic transition temperature. Below 30 K, $\rho_{//}$ rapidly increases to be a 100-times larger value from 30 down to 4 K, which was reported previously [11,14]. $\rho_{\perp}$ shows similar temperature dependence to $\rho_{//}$, but the magnitude is about 10 times larger, suggesting that the c-axis network of the zigzag Ir$_3$O$_{12}$ trimer is predominant in BaIrO$_3$. While $\rho_{\perp}$ is consistent with the result reported by Cao *et al*., our data did not reproduce their $\rho_{//}$ showing two anomalies near 26 and 80 K [11]. We should note here that we measured $\rho_{//}$ for three different single crystals, but did not observe any anomaly near 26 and 80 K, which is consistent with the susceptibility in Fig. 3(a).

Figure 3(c) shows the Seebeck coefficient $S_{//}$ and $S_{\perp}$ for the temperature gradient parallel and perpendicular to the c-axis, respectively. $S_{//}$ slightly increases with decreasing temperature above $T_C$, and is suddenly enhanced below $T_C$ to achieve 120 μV/K at 120 K. The sudden increase of $S_{//}$ below $T_C$ is due to the gap opening in the density of states which is consistent with the hump-type anomaly in $\rho_{//}$. Below 50 K, $S_{//}$ is roughly linear in temperature $T$ like a metal in contrast to the sharp increase of the non-metallic $\rho_{//}$. A preliminary data for $S_\perp$ (including larger errors than $S_{//}$ owing to the sample shape) exhibits a similar enhancement due to the gap opening just below $T_C$ and linear temperature dependence below 100 K. A similar temperature dependence of $S$ on a polycrystalline sample was observed by Kini *et al.* [14]. Since the Seebeck coefficient is a good probe for entropy per carrier, the $T$-linear $S_{//}$ and $S_\perp$ naturally indicate that the density of states is constant at low temperatures, suggesting no electronic phase transitions. Thus the rapid increase in $\rho_{//}$ and $\rho_\perp$ is unlikely to come from phase transitions.

A most striking feature of the title compound is anomalously giant nonlinear conduction at low temperatures. Cao et al. [11] first found the nonlinear conduction in BaIrO$_3$, which they ascribed to the sliding motion of CDW. Comparing with their data, we measured much more systematically as functions of current and temperature for several samples. We employed pulsed currents, which successfully excluded the heating effects, while they did not state how to avoid heating. Figure 4 shows the voltage-current characteristics as a function of pulsed current at several temperatures. The electric field $E$ at 4.2 K increases rapidly with increasing current density up to 2 A/cm$^2$, above which it decreases abruptly and shows a negative derivative resistance ($dE / dJ < 0$) up to 6 A/cm$^2$. Eventually, it becomes ohmic above 6 A/cm$^2$. A similar current dependence at 4.2 K was observed for $J$ perpendicular to the c-axis (not shown). This giant nonlinear conduction with the "mountain"-type anomaly is suppressed with increasing temperature, and almost disappears above 20 K. The sharp increase in resistivity for 0.01 A/cm$^2$ is suppressed by a current density of 20 A/cm$^2$, and the resistivity looks metallic down to 4 K for 20 A/cm$^2$ as shown in the inset of Fig. 4.

The $E$-$J$ characteristics are qualitatively similar to those of a thyristor [24]. As is well known, the thyristor is an electronic component consisting of two pairs of *p-n* diodes, and the nonlinear conduction is based on the interface effects at the *p-n* junctions. In contrast, the nonlinear conduction of BaIrO$_3$ is a *bulk* effect: This is observed in the four-probe configuration where the voltage drop is far from the interface at the current contact. We further note that the characteristic field (0.12 V/cm at 4.2 K) is extremely low, which clearly excludes the possibility for the hot electron and/or impact ionization. In such systems, a nonlinear response may occur, when the electric energy gained by the applied electric field $E$ exceeds the thermal energy $k_B T$. Then a characteristic length scale $L$ can be defined by $eEL = k_B T$. $L$ is estimated to be 30 μm at 4.2 K for 0.12 V/cm, which is much longer than the electron mean free path. Thus some other physics is required to explain the observed nonlinear conduction.

We do not think that the sliding motion of CDW is an origin for the nonlinear conduction in BaIrO$_3$ because of the following reasons: (i) the giant nonlinear conduction occurs only below 20 K, well below $T_C$=180 K. As already mentioned, no trace of phase transitions is detected in the susceptibility and the specific heat below 180 K. A nonlinear conduction would be seen for a CDW conductor at any temperatures below $T_C$, which implies that it is not related to the CDW transition. (ii) The nonlinear conduction was observed along both the directions parallel and perpendicular to the c axis. A CDW conductor would show a nonlinear conduction only along the nesting vector, and would retain the other two directions to be ohmic. (iii) The *E-J* characteristics are different from that of a CDW conductor. The latter shows an abrupt jump in resistivity at a certain threshold electric field, and rarely shows such a clear negative derivative resistance [29].

Although we have no clear answer for the mechanism of the giant nonlinear conduction in BaIrO$_3$, we can find many similarities to the *E-J* characteristics in the organic salt $\theta$-(BEDT-TTF)$_2$CsM(SCN)$_4$ [23, 24]. In fact, the two systems commonly share the above-mentioned features (i) (ii) and (iii). Competing charge orders play an important role in the organic salt, in which an external current melts charge-ordered domains of one particular modulation [24]. We expect essentially the same mechanism in BaIrO$_3$. Recent μSR measurement indicates that the existence of the local magnetic field of short range order at low temperatures [30], which could be associated with competing orders and the nonlinear conduction. Careful diffraction study with external currents is of course indispensable, which is now in progress.

In summary, single-crystalline samples of monoclinic BaIrO$_3$ were synthesized using a flux method, and their magnetization, resistivity, Seebeck coefficient and nonlinear voltage-current characteristics were measured. The magnetizations with a field parallel and perpendicular to the c axis show a rapid increase due to a ferromagnetic ordering below $T_C$ of 180 K. The resistivity along the c axis shows a nonmetallic conduction above $T_C$ and a metallic behavior from 120 down to 30 K, through a hump-type anomaly corresponding to a gap opening in density of states just below $T_C$. Finally it exhibits a sharp increase below 30K, which is highly nonlinear for external currents. The Seebeck coefficient suddenly increases below $T_C$, and is in proportion to temperature like a metal below 50 K. The anomalous giant nonlinear conduction is observed below 20 K, and the metallic conduction is induced with disappearance of the sharp increase below 30 K by a current density of 20 A/cm$^2$. A quantitative comparison with the nonlinear conduction of charge density wave has revealed that the giant nonlinear conduction in BaIrO$_3$ would belong to a novel class of collective excitation in solids.

**FIGURE CAPTOPNS:**

FIG. 1

Crystal structure of monoclinic $BaIrO_3$.

FIG. 2

(a) X-ray diffraction pattern of a powdered $BaIrO_3$ single-crystal at room temperature with the reflection positions calculated from the parameters reported by Siegrist [19]. (b) X-ray diffraction image of a $BaIrO_3$ single crystal for a certain direction.

FIG. 3

Temperature dependence of (a) magnetization with applying magnetic field parallel and perpendicular to the c axis, (b) resistivity with external current parallel and perpendicular to the c axis, and (c) Seebeck coefficient with temperature gradient parallel and perpendicular to the c axis.

FIG. 4

Non-linear voltage-current characteristics with pulsed current (pulse width = 12 ms) along the c axis at the several temperatures. Dashed line is the guide to eye. Inset: The temperature dependence of resistivity for $J = 0.02$ (the same $J$ for the data in Fig. 3(b)) and 20 $A/cm^2$ along the c axis.

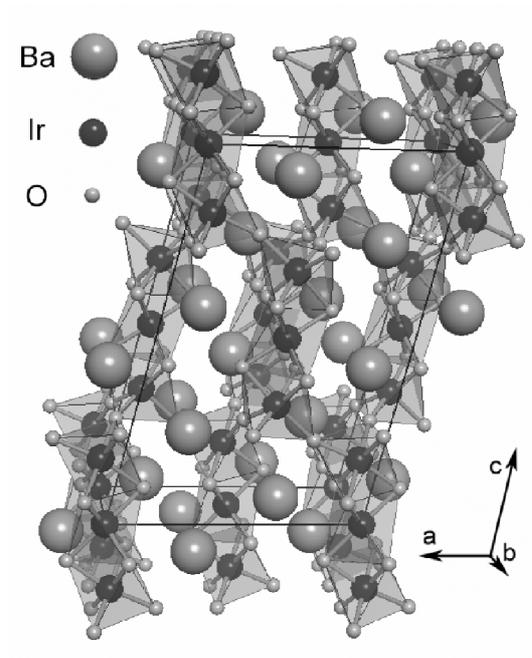

Figure 1:

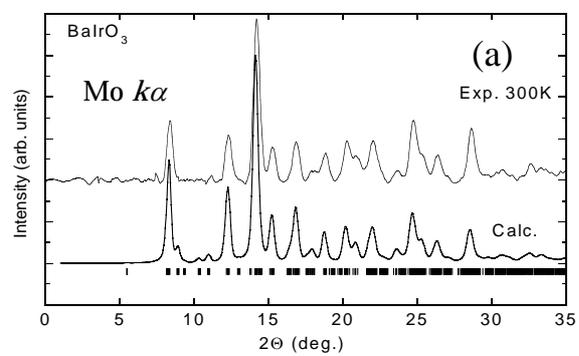

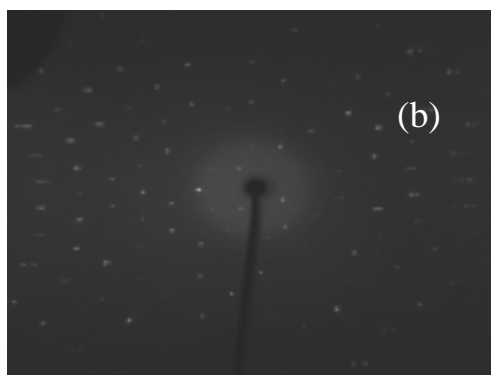

Figure 2:

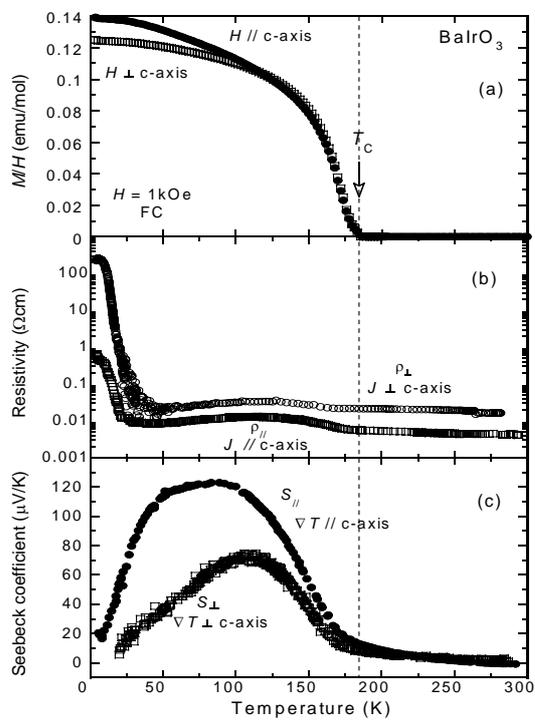

Figure 3:

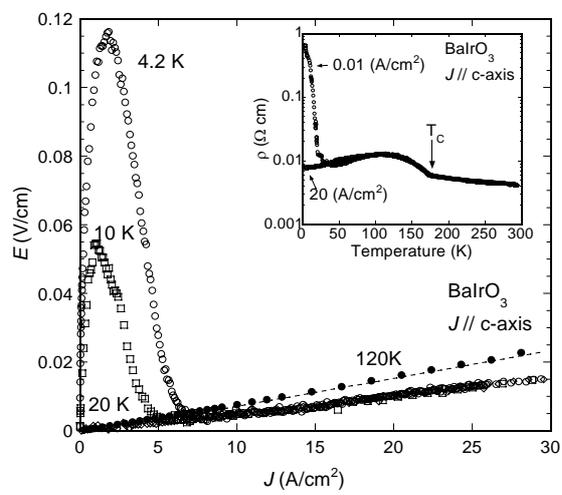

Figure 4: